\documentstyle[prb,aps,multicols]{revtex}
\begin{document}

\draft

\title{
Coherent control of macroscopic quantum states in a single-Cooper-pair box
}

\author{
Y. Nakamura$^*$, Yu.\ A. Pashkin$^{\dag}$ \& J. S. Tsai$^*$
}
\address{
$^*$NEC Fundamental Research Laboratories, Tsukuba, Ibaraki 305-8501, Japan\\
$^{\dag}$CREST, Japan Science and Technology Corporation (JST), Kawaguchi, Saitama 332-0012, Japan
}
\date{\today}
\maketitle

\begin{multicols}{2}

{\bf
A small superconducting electrode (a single-Cooper-pair box) connected to a reservoir via a Josephson junction constitutes an artificial two-level system, in which two charge states that differ by $2e$ are coupled by tunneling of Cooper pairs.
Despite its macroscopic nature involving a large number of electrons, the two-level system shows coherent superposition of the two charge states \cite{Bouchiat98,Joyez94,Flees97,Nakamura97}, and has been suggested as a candidate for a qubit, i.e.\ a basic component of a quantum computer\cite{Shnirman97,Averin98,Makhlin98}.
Here we report on time-domain observation of the coherent quantum-state evolution in the two-level system by applying a short voltage pulse that modifies the energies of the two levels nonadiabatically to control the coherent evolution.
The resulting state was probed by a tunneling current through an additional probe junction.
Our results demonstrate coherent operation and measurement of a quantum state of a single two-level system, i.e.\ a qubit, in a solid-state electronic device.
}

Rapidly improving nanofabrication technologies have made quantum two-level systems in solid-state devices promising for functional quantum circuit integration.
To coherently control an individual two-level system as a unit of such circuits, several systems have been examined, such as electronic\cite{Schedelbeck97,Oosterkamp98,Bonadeo98} and spin\cite{Loss98} states in quantum dots, nuclear spins of impurity atoms embedded in a substrate\cite{Kane98}, and magnetic-flux states in a superconducting ring\cite{Leggett84,Tesche90}.
However, only optical coherent control has been realized experimentally\cite{Bonadeo98}.

A single-Cooper-pair box\cite{Bouchiat98} (Fig.~1) is a unique artificial solid-state system in the sense that: 
(i) Although there is a large number of electrons in the metal `box' electrode, under superconductivity they all form Cooper pairs and condense into a single macroscopic ground state, $|n\rangle$, separated by a superconductivity gap $\Delta$ from the excited states with quasiparticles. Here $|n\rangle$ denotes the charge-number state with the excess number of electrons in the box, $n$. (ii) The only low-energy excitations are the transitions between different $|n\rangle$ states due to Cooper-pair tunneling through the Josephson junction, if $\Delta$ is larger than the single-electron charging energy of the box $E_C$. (iii) $E_C$, if larger than the Josephson energy $E_J$ and the thermal energy $k_BT$, suppresses a large fluctuation of $n$. Hence, we can consider the system an effective two-level system by taking into account the two lowest-energy states which differ by one Cooper pair. (iv) In addition, the relative energy of the two levels can be controlled through the gate voltage.
For example, as shown in Fig.~2a, electrostatic energies, $E_C(n-Q_t/e)^2$, of two such charge states $|0\rangle$ and $|2\rangle$, change as a function of the total gate-induced charge $Q_t$ and cross each other at $Q_t/e=1$. (The parabolic background energy is subtracted.)
In the presence of Josephson coupling, and with weak enough dissipation\cite{Neumann94}, these charge states would be coherently superposed and form two anticrossing eigenenergy bands (dashed curves in Fig.~2a).
The existence of the coherence has been inferred in energy-domain experiments by measuring ground-state properties\cite{Bouchiat98,Joyez94} and by spectroscopy\cite{Flees97,Nakamura97}.
However, coherent control and observation of quantum-state evolution in the time domain has not been achieved.
Such time-domain techniques are necessary to enable applications based on quantum coherent evolution\cite{Shnirman97,Averin98,Makhlin98}.

To investigate the coherent evolution, we applied a sharp voltage pulse to the pulse gate to control energy levels of the charge states and to manipulate the quantum state as shown in Figs.~2a and b.
If we select an initial condition $Q_t=Q_0$ far to the left from the resonance point (where $Q_0$ is the dc-gate induced charge), the initial state would, with a large probability $\sim 1$, be the ground state which is almost the pure $|0\rangle$ state.
The pulse brings the two charge states into resonance and lets the wave function coherently evolve between $|0\rangle$ and $|2\rangle$ during the pulse length $\Delta t$.
The quantum state at the end of the pulse would be a superposition of the two charge states which depends on $\Delta t$.
Here, the rise/fall time of the pulse must be short compared to the coherent oscillation time $h/E_J$, otherwise the state just follows the ground-state energy band adiabatically.

The probe junction was voltage-biased with an appropriate voltage $V_b$ so that $|2\rangle$ decays to $|0\rangle$ with two sequential quasiparticle tunneling events through the probe junction with predictable rates $\Gamma_{\rm qp1}$ and $\Gamma_{\rm qp2}$ (about $(6$~ns)$^{-1}$ and $(8$~ns)$^{-1}$ in the present experiment), where the latter state $|0\rangle$ is stable against the quasiparticle tunneling\cite{Nakamura97}.
The role of the quasiparticle tunneling is twofold.
One is the detection of $|2\rangle$ as two tunneling electrons. 
Since this ``detector" is always connected to the two-level system even during the pulse, a large probe junction resistance $R_b$ is necessary for small $\Gamma_{\rm qp1}$ ($\propto R_b^{-1}$) to avoid excessively disturb the coherence.
The other role is the preparation of the initial state for the next pulse operation by relaxation to the ground state.
With an array of pulses with a repetition time $T_r$ longer than the relaxation time, we can repeat the pulse operation many times and measure the dc current through the probe junction which would reflect the population in $|2\rangle$ after each pulse operation.

In the experiment, the actual pulse height at the pulse gate was not measurable, so we swept the range of dc-induced charge $Q_0$, with a fixed pulse height and the repetition time $T_r$.
Figure~2c shows the current through the probe junction versus $Q_0$.
Without a pulse array (dashed line), a broad current peak appeared at $Q_0/e=1$ where charge states $|0\rangle$ and $|2\rangle$ are degenerate.
This current is the Josephson-quasiparticle (JQP) current\cite{Fulton89,Averin89} and is carried by a cyclic process consisting of one Cooper-pair tunneling between the two charge states and two sequential quasiparticle tunneling events at the probe junction.
When applying a pulse array (solid line), on the left side of the JQP peak we observed a pulse-induced current with several peaks whose positions did not depend on $T_r$ but on $\Delta t$.
In Fig.~3a we extract the pulse-induced part of the current, $\Delta I$, for the pulse length $80\leq \Delta t \leq 450$~ps.
With increasing $\Delta t$ all the peaks moved towards smaller $Q_0$ and disappeared at $Q_0/e\sim 0.5$.
The region where the peaks existed extended to smaller $Q_0$ linearly with increasing pulse height (data not shown).

To compare with the experimental results, we simulated the pulse operation of the quantum state by numerically solving a time-dependent Schr\"odinger equation.
We calculated the average increase in the probability density at $|2\rangle$ after a single-pulse operation, $\langle\Delta P(2)\rangle$, which should approximately be proportional to $\Delta I$.
To adjust the maximum oscillation period in the time domain, $T_{\rm coh}$, we used Josephson energy $E_J=51.8$~$\mu$eV, and to adjust $Q_0$($=0.51e$) where $T_{\rm coh}$ was observed we used an effective pulse height $\Delta Q_p/e$ of 0.49.
The overall features of the pulse-induced current were reproduced (Fig.~3b).
$Q_0$ where $T_{\rm coh}$ was observed corresponded to the point where the applied pulse brought the two levels into resonance and $T_{\rm coh}$ equaled $h/E_J$.
The oscillation period in the time domain changed according to $h/\sqrt{E_J^2+\delta E^2}$, where $\delta E\equiv 4E_C \{ (Q_0+\Delta Q_p)/e-1 \}$ is the electrostatic energy difference between the two charge states during the pulse.

Figure~4 shows the pulse-induced current at $Q_0/e=0.51$ as a function of $\Delta t$, demonstrating that the coherent oscillation can be observed in the time domain and that we can control the quantum state through an arbitrary pulse length $\Delta t$.
The oscillation amplitude was smaller than that simply expected from $2e$ per pulse, $2e/T_r=20$~pA.
The finite rise/fall times of the pulse might explain this deviation.
Recall that in the long-rise/fall-time limit (adiabatic limit) there would be no transition probability to $|2\rangle$.
For the realistic rise/fall times of the pulse we assumed in the simulation above, for example, the amplitude of the oscillations in $\langle\Delta P(2)\rangle$ at $Q_0/e=0.51$ is reduced to $\sim 0.4$, by which the current signal would be decreased.
Moreover, the finite repetition time not much longer than $\Gamma_{\rm qp1}^{-1}+\Gamma_{\rm qp2}^{-1}$ could also reduce the signal due to the incomplete relaxation of $|2\rangle$ to $|0\rangle$ after each pulse.

To further confirm that the observed oscillation was coherent oscillation due to Josephson coupling, we estimated Josephson energy $E_J$ from the oscillation period $T_{\rm coh}$ as $E_J=h/T_{\rm coh}$ and investigated its magnetic-field dependence (black dots in the inset of Fig.~4).
We also measured $E_J$ in the frequency domain through microwave spectroscopy of the energy-level splitting\cite{Nakamura97} (open squares).
Those data agreed very well with each other and fit the expected cosine curve.

For future application as quantum computing devices\cite{Shnirman97,Averin98,Makhlin98}, a crucial parameter is the decoherence time during which coherent evolution is maintained.
The main decoherence source in a single-Cooper-pair box is thought to be spontaneous photon emission to the electromagnetic environment\cite{Bouchiat98,Shnirman97,Averin98,Makhlin98}, and the decoherence time could exceed 1~$\mu$s.
With a probe junction, though as in our setup, the ``detection" with quasiparticle tunneling through the probe junction would be the main source of decoherence.
So far, we have observed oscillation up to $\Delta t\sim 2$~ns, although low-frequency background-charge fluctuation degraded the dc current signal and made it difficult to determine the envelope of the decay.
A more detailed study of the decoherence time would provide important information for designing solid-state quantum circuits using superconducting single-Cooper-pair boxes.

\acknowledgements

We thank W. Hattori, M. Baba and H. Suzuki for experimental help and M. Ueda, Y. Kohno, M. H. Devoret and Y. Ootuka for valuable discussions. This work has been supported by the Core Research for Evolutional Science and Technology (CREST) of the Japan Science and Technology Corporation (JST).

Correspondence and requests for materials should be addressed to Y. N. ($e$-mail: yasunobu@frl.cl.nec.co.jp).

\pagebreak

\narrowtext

\begin{figure}
\caption{
Single-Cooper-pair box with a probe junction.
{\bf a}, Micrograph of the sample. 
The electrodes were fabricated by electron-beam lithography and shadow evaporation of Al on a SiN$_x$ insulating layer (400-nm thick) above a gold ground plane (100-nm thick) on the oxidized Si substrate.
The `box' electrode is a $700\times 50\times 15$~nm$$ Al strip containing $\sim10^8$ conduction electrons.
The reservoir electrode was evaporated after a slight oxidation of the surface of the box so that the overlapping area becomes two parallel low-resistive tunnel junctions ($\sim 10$~k$\Omega$ in total) with Josephson energy $E_J$ which can be tuned through magnetic flux $\phi$ penetrating through the loop.
Before the evaporation of the probe electrode we further oxidized the box to create a highly-resistive probe junction ($R_b\sim 30$~M$\Omega$).
Two gate electrodes (dc and pulse) are capacitively coupled to the box electrode.
The sample was placed in a shielded copper case at the base temperature ($T\sim 30$~mK; $k_BT\sim 3$~$\mu$eV) of a dilution refrigerator.
The single-electron charging energy of the box electrode $E_C\equiv e^2/2C_\Sigma$ was $117\pm 3$~$\mu$eV, where $C_\Sigma$ is the total capacitance of the box electrode.
The superconducting gap energy $\Delta$ was $230\pm 10$~$\mu$eV.
{\bf b}, Circuit diagram of the device.
The $C$'s represent the capacitance of each element and the $V$'s are the voltage applied to each electrode.
}
\end{figure}

\begin{figure}
\caption{
Pulse modulation of quantum states.
{\bf a}, Energy diagram illustrating electrostatic energies (solid lines) of two charge states $|0\rangle$ and $|2\rangle$ (with the number of excess charges in the box $n=0$ and 2) as a function of the total gate-induced charge $Q_t\equiv Q_0+C_pV_p(t)$, where $Q_0\equiv C_gV_g+C_bV_b$ is the dc-gate induced charge.
The dashed curves show eigenenergies (in the absence of the quasiparticle tunneling at the probe junction) as a function of $Q_t$.
Suppose that before a pulse occurs, $Q_t$ equals $Q_0$, which is far from the resonance point, and the system is approximately in the pure charge state $|0\rangle$ (black dot).
Then, a voltage pulse of an appropriate height abruptly brings the system into resonance $Q_t/e=1$ (solid arrow), and the state starts to oscillate between the two charge states.
At the end of the pulse, the system returns to $Q_t=Q_0$ (dashed arrow) with a final state corresponding to the result of the time evolution.
Finally, the $|2\rangle$ state decays to $|0\rangle$ with two quasiparticle tunneling events through the probe junction with rates of $\Gamma_{\rm qp1}$ and $\Gamma_{\rm qp2}$ (dotted arrows).
{\bf b}, Schematic pulse shape with a nominal pulse length $\Delta t$ (solid line).
The rise/fall times of the actual voltage pulse was about 30--40~ps at the top of the cryostat.
The voltage pulse was transmitted through a silver-plated Be-Cu coaxial cable (above 4.2~K), a Nb coaxial cable (below 4.2~K) and an on-chip coplanar line to the open-ended pulse gate shown in Fig.~1a.
The insets illustrate situations of the energy levels before/during/after the pulse.
{\bf c}, Current through the probe junction versus $Q_0$ with (solid line) and without (dashed line) the pulse array.
The pulse length was $\Delta t=160$~ps and the repetition time was $T_r=16$~ns.
The data were taken at $V_b=650$~$\mu$V and $\phi/\phi_0=0.31$, where $\phi_0\equiv h/2e$ is a flux quantum.
}
\end{figure}

\begin{figure}
\caption{
Effect of pulsation as a function of dc-induced charge $Q_0$ and pulse length $\Delta t$.
{\bf a}, 3D plot of pulse-induced current $\Delta I$ which is the difference between currents measured with and without a pulse array.
$\Delta t=80$~ps was the shortest pulse length available with our pulse-pattern generator (Anritsu MP1758A).
{\bf b}, Calculated average increase in probability density at $|2\rangle$ after a single-pulse operation, $\langle \Delta P(2)\rangle$.
The averaged probability density after the pulse was calculated by numerically solving a time-dependent Schr\"odinger equation and by averaging out small residual oscillations in the time domain.
The effect of decoherence was not included.
As the initial condition of the Schr\"odinger equation, we used a mixture of two eigenstates at $Q_t=Q_0$ with weights obtained from a steady-state solution of density-matrix equations that describe charge transport through the device in the absence of a pulse array.
The initial probability density was also calculated from the steady-state solution.
In the calculations, Josephson energy $E_J=51.8$~$\mu$eV and an effective pulse height $\Delta Q_p/e=0.49$ were used.
The solid line in Fig.~2b shows an example (at $\Delta t=300$~ps) of the pulse shape used in this calculation.
}
\end{figure}

\begin{figure}
\caption{
Pulse-induced current as a function of pulse length $\Delta t$.
The data correspond to the cross-section of Fig.~3a at $Q_0/e=0.51$.
The inset shows Josephson energy $E_J$ versus the magnetic flux $\phi$ penetrating through the loop.
$E_J$ was estimated by two independent methods.
One was from the period of the coherent oscillation $T_{\rm coh}$ as $h/T_{\rm coh}$.
The other was from the gap energy observed in microwave spectroscopy\protect\cite{Nakamura97}.
The solid line shows a fitting curve with $E_J(\phi =0)=84$~$\mu$eV assuming cosine $\phi$-dependence of $E_J$.
}
\end{figure}

\end{multicols}

\end{document}